\documentclass[sigconf,screen,pbalance]{acmart}
  
  \AtBeginDocument{%
    }

  \usepackage{algorithm}
  \usepackage{algpseudocodex}
  \usepackage{graphicx}
  \usepackage{textcomp}
  \usepackage{xcolor}
  \usepackage{float}
  \usepackage{subcaption}
  \usepackage{booktabs}
  \usepackage{amsmath}
  \usepackage{multirow}
  \usepackage{colortbl}
  \definecolor{gold}{HTML}{FFD700}
  \definecolor{silver}{HTML}{C0C0C0}
  \definecolor{bronze}{HTML}{CD7F32}
  \usepackage{tikz}
  \usetikzlibrary{arrows.meta,calc,fit,positioning}
  
  \usepackage{booktabs}
  \usepackage{enumitem}
  \usepackage{hyperref}
  \usepackage[para]{footmisc}
  \usepackage{pifont}
  \usepackage{listings}
  \newcommand{\observation}[3]{%
    \par
    \noindent\begingroup
    \setlength{\fboxrule}{0.9pt}%
    \setlength{\fboxsep}{5pt}%
    \fcolorbox{black}{black!3}{%
      \parbox{\dimexpr\linewidth-2\fboxsep-2\fboxrule\relax}{%
        \textbf{Observation #1. #2}\par
        #3%
      }%
    }%
    \endgroup
    \par
  }
  \algnewcommand{\IfReturn}[2]{%
    \If{#1} \textbf{return} #2; \EndIf%
  }
  \setcopyright{acmlicensed}
  \copyrightyear{2026}
  \acmYear{2026}
  \acmDOI{10.1145/3848591.3856833}
  \acmConference[RACS '26]{International Conference on Resilience, AI and Cyber-Systems}{Nov. 17--20,
    2026}{Fukuoka, Japan}
  \acmISBN{XXX-X-XXXX-XXXX-X/2026/11}

\begin{document}
  \title[]{Towards Training Private LLMs: Exploring Fine-Tuning Language Models on Apple Silicon with RDMA over Thunderbolt}
  \titlenote{This paper has been accepted by RACS '26.}


  \author{En-Ming Huang$^{1}$, Yao-Ting Hsieh$^{2\dagger}$, Hsiang-Yu Tsou$^{1}$, Mu-Chi Chen$^{1\dagger}$,\\
  Shih-Hao Hung$^{1\dagger}$ and H.T. Kung$^{3\dagger}$}
  \affiliation{%
    \institution{$^{1}$National Taiwan University, $^{2}$Academia Sinica, $^{3}$Harvard University, {$^{\dagger}$}SiliconMind}
    \city{}
    \country{}}
  


  \renewcommand{\shortauthors}{Huang et al.}

  \begin{abstract}
Private large language model (LLM) fine-tuning is increasingly important for organizations that need to adapt models using sensitive data, but it often exceeds the memory capacity of commodity datacenter accelerators. Apple Silicon offers a different design point through large unified memory and lower complete-system cost, while recent Apple software support enables distributed execution over RDMA-over-Thunderbolt (TB). This paper studies whether Apple Silicon can serve as a practical platform for private LLM fine-tuning. We characterize RDMA-over-TB communication on Mac Studio nodes, showing that the measured bandwidth is far below nominal TB specifications. Next, we extend Apple's implementation with multi-trunk communication, persistent worker threads, and CPU-side gradient overlap to better exploit multiple direct TB links for LLM fine-tuning workloads. Finally, on a four-node Mac Studio cluster that fine-tunes a Qwen3-9B, our optimizations improve weak-scaling throughput by up to \(1.6\times\) over the single-trunk, non-overlapped baseline and reach 936 tokens/s for sequence length 17408. We further compare Apple Silicon with an NVIDIA H100 platform to quantify the trade-off between memory capacity, throughput, and acquisition cost, showing that Apple Silicon can provide a cost-effective solution for private LLM fine-tuning.
\end{abstract}

  \keywords{}

  \maketitle

  \section{Introduction}
\label{sec:intro}

Private adaptation of large language models (LLMs) is becoming increasingly important for organizations that need to incorporate proprietary knowledge without exposing sensitive data to external services~\cite{hanke2024openllm}. Internal documents, source code repositories, and domain-specific design artifacts often contain information that cannot be sent to hosted APIs, yet these data are precisely what make task-specific fine-tuning valuable. Recent work has explored private LLM deployment on Apple Silicon platforms~\cite{chen2025privateapplellm}, and demonstrated the effectiveness of model fine-tuning for specialized and sensitive workflows such as automated hardware design~\cite{siliconmind,QiMeng2025CodeVR1}.

Compared with pre-training, fine-tuning requires substantially fewer training tokens, but memory capacity remains a critical challenge, particularly for long-context workloads. Modern reasoning and agentic applications frequently process long execution traces, tool outputs, and generated candidates whose context cannot be trivially partitioned without altering task semantics~\cite{siliconmind, EvolVE, yao2023react}. During training, activations, gradients, and model parameters must be retained for backpropagation, causing memory requirements to grow rapidly with sequence length; optimizers further increase this footprint by maintaining optimizer states~\cite{zero2020,kingma2014adam,korthikanti2023reducing}. Consequently, systems with large amounts of accelerator-accessible memory are attractive platforms for private model adaptation.

\begin{figure}[t]
  \centering
  \includegraphics[width=0.9\linewidth]{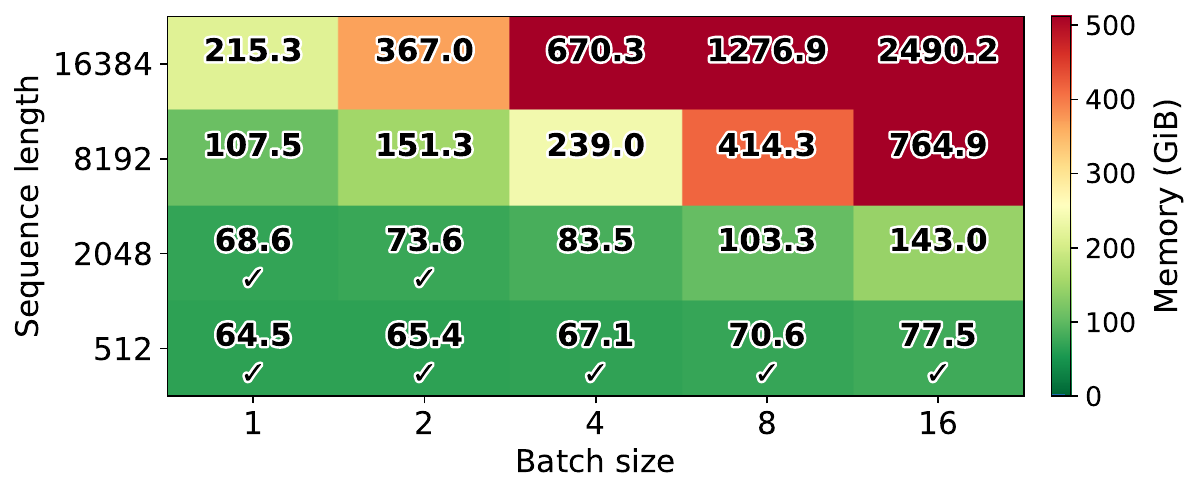}
  \vspace{-1.5em}
  \caption{Theoretical minimum memory requirement for Qwen3-9B~\cite{yang2025qwen3technicalreport} fine-tuning in FP32 with SGD.}
  \vspace{-1.5em}
  \label{fig:mem-temp}
\end{figure}

Apple Silicon occupies a unique position in this design space due to its large system memory and lower cost. A single M3 Ultra system provides up to 512~GiB of unified memory that is directly accessible by the integrated GPU, significantly exceeding the memory capacity of most commodity accelerator platforms (e.g., an NVIDIA H100 provides only 80~GiB of memory)~\cite{appleMacStudioSpecs}. At the same time, an entire M3 Ultra system costs about \$10{,}000, compared to approximately \$30{,}000 for a single H100 accelerator and around \$300{,}000 for a complete 8$\times$H100 DGX workstation.
Fig.~\ref{fig:mem-temp} illustrates this memory gap concretely using the theoretical minimum memory requirement for training Qwen3-9B~\cite{yang2025qwen3technicalreport} in FP32 with stochastic gradient descent (SGD), which does not require optimizer states. While a sequence length of 512 requires 65--78~GiB across the shown batch sizes, the requirement grows rapidly with context length and batch size, reaching 215~GiB at a sequence length of 16{,}384 with a batch size of one. This exceeds the capacity of a single H100 for longer contexts, whereas several configurations still fit within the 512~GiB unified memory of a single Mac Studio node.

Concurrently, recent software advancements through the Apple-developed MLX framework~\cite{mlx2023} enable distributed training across multiple nodes and introduce \textit{JACCL}, a communication backend that supports Remote Direct Memory Access (RDMA) over Thunderbolt (TB)~\cite{mlxDistributed}. These developments render multi-node Apple Silicon clusters a compelling candidate for distributed fine-tuning; however, the practical communication performance of TB-based RDMA for LLM training workloads remains largely uncharacterized.

Unlike datacenter systems that rely on NVLink or InfiniBand, Apple Silicon nodes communicate over TB and Ethernet. Although TB provides substantial raw bandwidth, it is exposed as point-to-point links whose effectiveness depends on topology, collective implementation, and software concurrency. To understand these constraints, we conduct a systematic study of distributed LLM fine-tuning on Apple Silicon. We first characterize and optimize TB-based RDMA communication in MLX through link trunking and multi-threaded communication. We then evaluate collective communication performance and show that trunking together with CPU-side gradient overlap can substantially improve model training on Apple Silicon. Finally, we perform weak-scaling experiments using supervised fine-tuning of the Qwen3-9B model~\cite{yang2025qwen3technicalreport} on up to four M3 Ultra nodes and compare with H100.

This paper makes the following contributions.
\begin{itemize}[]
\item We characterize RDMA-over-TB communication on Apple Silicon for LLM training workloads and concludes with four observations. Our microbenchmarks quantify latency, bandwidth, all-reduce performance, and multi-node scaling.

\item We extend the \textit{JACCL} backend with multi-trunk RDMA communication and persistent worker threads, allowing one logical peer transfer to use multiple TB links in parallel. We also introduce CPU-side gradient overlap for MLX training. Together, these optimizations improve end-to-end weak-scaling throughput by up to \(1.6\times\) over the single-trunk, non-overlapped \textit{JACCL} baseline.

\item We evaluate Qwen3-9B supervised fine-tuning on Apple Silicon from single-node and distributed perspectives. The results show that a 512~GB unified-memory node can fit long-context configurations without context parallelism~\cite{li2023sequence}, and that a four-node Mac Studio cluster achieves efficient weak scaling with multi-trunk communication and overlap. We further compare system-level performance, memory capacity, and cost against an NVIDIA H100 platform.
\end{itemize}

The rest of this paper is organized as follows. Section~\ref{sec:background} reviews the background. Section~\ref{sec:methodology} presents the communication optimizations. Section~\ref{sec:sft} evaluates model training performance against H100. Section~\ref{sec:conclusion} concludes.

  \section{Background}
\label{sec:background}

This section provides the background needed to understand distributed fine-tuning on Apple Silicon. We first describe the unified-memory platform and RDMA-over-TB interface that make Mac Studio clusters possible, then explain how MLX and \textit{JACCL} expose this interconnect to distributed applications. We next contrast communication patterns in LLM inference and training to motivate our focus on data-parallel fine-tuning, and finally summarize the memory states that make long-context training capacity-intensive.

\subsection{Apple Silicon and RDMA over TB}

Apple Silicon integrates CPU, GPU, and memory into a unified-memory system. This design differs from conventional GPU servers, where GPU memory is separated from host memory and high-end interconnects such as NVLink or InfiniBand are used to connect accelerators. In our target platform, each Mac Studio provides large Low-Power Double Data Rate 5 (LPDDR5) memory capacity and six TB5 ports, but the practical node-to-node interconnect choices are limited to built-in 10 Gigabit Ethernet (10GbE) and TB5. Table~\ref{tab:apple-silicon-interconnects} summarizes these interfaces and their tradeoffs; although 10GbE is available on the platform, this work focuses on RDMA over TB because it offers lower latency, while its point-to-point links motivate the trunking design studied later.

\begin{table}[t]
  \centering
  \caption{Node-level interconnection options on Apple Silicon Mac Studio systems.}
  \label{tab:apple-silicon-interconnects}
  \small
  \setlength{\tabcolsep}{3pt}
  \begin{tabular}{@{}>{\raggedright\arraybackslash}p{0.20\columnwidth}>{\raggedright\arraybackslash}p{0.12\columnwidth}>{\raggedright\arraybackslash}p{0.25\columnwidth}>{\raggedright\arraybackslash}p{0.35\columnwidth}@{}}
    \toprule
    Interface & Count & Nominal bandwidth & Pros / cons \\
    \midrule
    TB5 RDMA & 6 ports & 80~Gb/s per direction; 160~Gb/s bidirectional total & \textbf{Pro}: lower latency. \textbf{Con}: point-to-point links only. \\
    10GbE & 1 port & 10~Gb/s Ethernet & \textbf{Pro}: conventional Ethernet connectivity. \textbf{Con}: lower bandwidth and higher latency\\
    \bottomrule
  \end{tabular}
\end{table}

Starting from macOS~26.2, Apple supports RDMA over TB on Apple Silicon Macs with TB5~\cite{appleRDMAThunderbolt,mlxDistributed}. This exposes an \textit{InfiniBand Verbs}-compatible interface for the TB controller. Developers can use the \textit{Verbs} API directly by including \texttt{infiniband/verbs.h} and linking against \texttt{librdma.tbd} from the macOS SDK~\cite{appleRDMAThunderbolt}. Applications open an RDMA device, allocate communication resources, create queue pairs, and post send or receive work requests.

Apple's RDMA over TB interface is narrower than a full datacenter RDMA stack. It supports send and receive operations only, with at most 10 Unreliable Connection (UC) queue pairs and at most 4095 outstanding work requests at a time~\cite{appleRDMAThunderbolt}. UC queue pairs lack the full reliability semantics of Reliable Connection (RC), which features data integrity checks and remote memory access. In our setup, each TB5 link is exposed as a \textit{Verbs}-accessible RDMA interface, so the six TB ports on a Mac Studio can be used as separate communication paths between machines.

\subsection{\textit{JACCL} with MLX}

Developed by Apple, MLX is an array framework similar to PyTorch that also provides distributed communication backends for multi-node execution, including a TCP-based ring backend and \textit{JACCL}, an RDMA-over-TB backend~\cite{mlxDistributed}. These two backends differ fundamentally in how peers are addressed. The ring backend relies on TCP sockets, allowing any node to communicate with any other node using only network addresses and ports, making it largely topology agnostic. In contrast, RDMA over TB is inherently point-to-point: each TB5 connection appears as an independent RDMA device, and communication requires explicit knowledge of which local TB device is connected to each remote peer.

\textit{JACCL} is implemented on top of InfiniBand \textit{Verbs} and uses a hostfile to specify the mapping between peer pairs and their corresponding RDMA devices. Consequently, \textit{JACCL} assumes a fully connected mesh topology in which every pair of nodes is connected by a dedicated TB cable. This assumption enables communication patterns that exploit direct pairwise connectivity. For collective operations such as all-reduce, \textit{JACCL} employs a broadcast-based algorithm in which each node directly exchanges data with all other nodes, eliminating the multi-hop communication stages required by ring-based implementations.

\begin{figure}[t]
  \centering
  \includegraphics[width=0.99\linewidth]{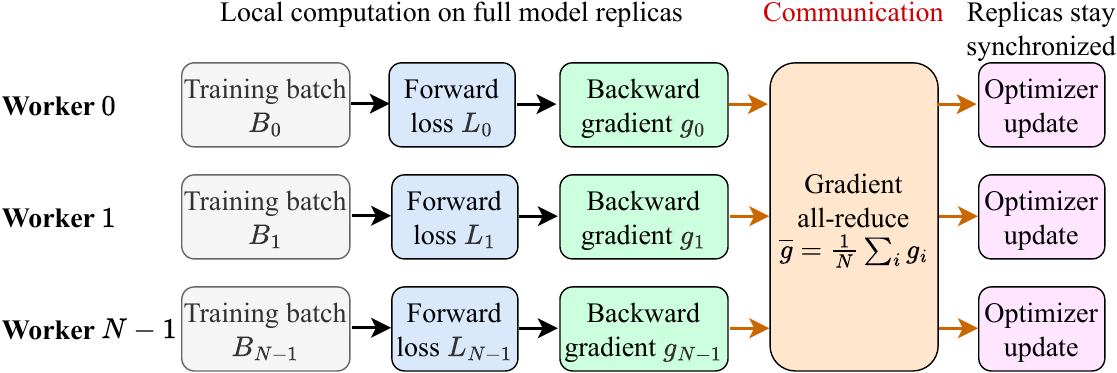}
  \vspace{-1em}
  \caption{Data-parallel LLM training flow. Each worker keeps a full model replica and processes a different mini-batch shard. Forward and backward passes are local computations; communication is concentrated in the gradient all-reduce before all workers apply the same optimizer update.}
  \label{fig:data-parallel-training}
  \Description{A schematic of data-parallel training showing local forward and backward computation on each worker, followed by gradient all-reduce communication and synchronized optimizer updates.}
  \vspace{-1em}
\end{figure}

\subsection{Communication in LLM Inference and Training}
\label{sec::comm-llm}

Distributed LLM inference and training exhibit fundamentally different communication characteristics. Inference is latency-sensitive because each autoregressive decoding step must complete before the next token can be generated. To reduce token latency, inference systems commonly employ tensor parallelism, which partitions matrix operations across devices and synchronizes partial results through collective operations~\cite{shoeybi2019megatron,narayanan2021megatron}. These collectives occur at every layer, making communication both frequent and latency critical. The communicated payload for a single hidden-state vector is also relatively small (e.g., 8--16~KiB in FP16 for hidden dimensions of 4096--8192~\cite{yang2025qwen3technicalreport,2025llama3}), limiting opportunities to amortize communication latency or multi-link scheduling overhead.

In contrast, LLM training is typically throughput-oriented. When memory capacity permits model replication across workers, data parallelism is attractive because communication is concentrated in gradient synchronization after backpropagation~\cite{narayanan2021megatron,2025llama3}. As illustrated in Fig.~\ref{fig:data-parallel-training}, forward and backward computations are performed locally on each worker, while communication occurs primarily in gradient all-reduce operations before synchronized optimizer updates. Large-scale deployments often combine data, tensor, pipeline, and context parallelism to overcome memory and scalability constraints, but these techniques introduce additional cross-device communication during both forward and backward execution~\cite{shoeybi2019megatron,narayanan2021megatron,2025llama3}.

This distinction motivates our focus on fine-tuning workloads. The small, latency-critical messages generated by tensor-parallel inference are unlikely to benefit from software trunking over RDMA-over-TB links. In contrast, the larger and more bandwidth-intensive gradient exchanges characteristic of data-parallel training provide a more suitable workload for evaluating whether TB link trunking can improve end-to-end training performance.

\subsection{Memory Requirements in Model Inference and Training}

LLM memory requirements are primarily driven by model size, context length, and the operations being performed. Table~\ref{tab:llm-memory-requirements} lists the main memory state in each mode. Inference memory consists of static model weights plus dynamic working memory. The key--value (KV) cache stores context history and scales with context length and concurrent requests; for long contexts such as \(128\text{K}\), it can reach tens of GiB per active request depending on model size and precision. During training, the same attention tensors are not kept as a persistent generation cache; they are part of the forward activations saved or recomputed for backpropagation. Training is heavier because it must track gradients, optimizer state, and activations. In full fine-tuning, these extra tensors often make training require roughly \(6\times\)--\(8\times\) more memory than inference~\cite{zero2020}.

\begin{table}[t]
  \centering
  \caption{Main memory state in LLM inference and training.}
  \vspace{-1.3em}
  \label{tab:llm-memory-requirements}
  \small
  \setlength{\tabcolsep}{3pt}
  \begin{tabular}{@{}>{\raggedright\arraybackslash}p{0.24\columnwidth}>{\raggedright\arraybackslash}p{0.3\columnwidth}>{\raggedright\arraybackslash}p{0.38\columnwidth}@{}}
    \toprule
    State & Inference & Training \\
    \midrule
    Model weights & Static weights & Weights to update \\
    KV cache & Context history & Part of activations \\
    Activations & Temporary buffers & Saved (or recomputed) \\
    Gradients & No & Per trainable parameter \\
    Optimizer state & No & Momentum/variance buffers \\
    \bottomrule
  \end{tabular}
\end{table}

  \section{Methodology}
\label{sec:methodology}

This section describes the communication optimizations used for distributed fine-tuning on Apple Silicon. We first introduce a multi-trunk RDMA-over-Thunderbolt design that extends the \textit{JACCL} backend to use multiple physical links between a pair of nodes. We then describe a layer-wise overlap mechanism that schedules CPU-side gradient synchronization concurrently with GPU backpropagation. Finally, we characterize latency, bandwidth, software concurrency, and multi-node scaling through microbenchmarks, including a comparison with the Ethernet-based MLX backend.

\begin{figure}[t]
    \centering
    \includegraphics[width=.99\linewidth]{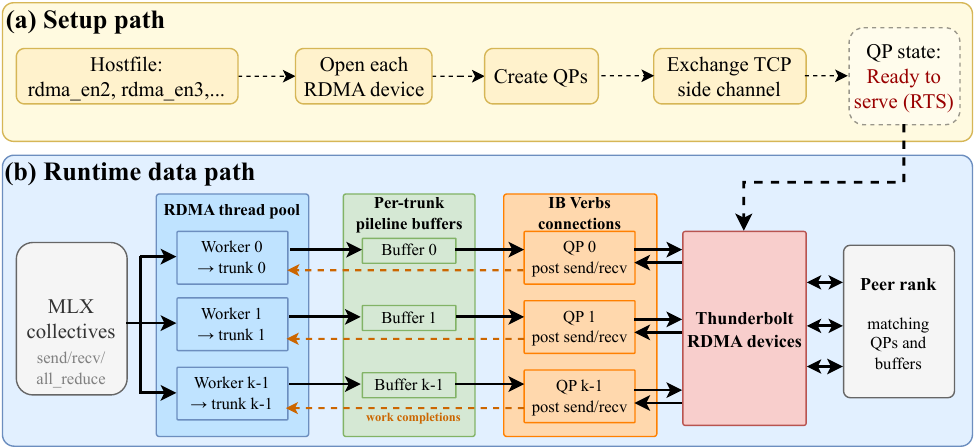}
    \vspace{-1.1em}
    \caption{Multi-trunk RDMA-over-Thunderbolt communication path. Persistent workers stripe each collective message across multiple TB RDMA links.}
    \Description{Block diagram showing a setup phase that maps hostfile RDMA interfaces to communication workers, followed by a runtime phase that chunks collective messages and stripes them across multiple Thunderbolt RDMA links.}
    \label{fig:flow}
\end{figure}

\subsection{Multi-trunk RDMA over TB}
\label{sec::multi-trunk}

Fig.~\ref{fig:flow} summarizes our communication path. A single TB connection exposes one RDMA device and one queue-pair stream, while each Mac Studio provides multiple independent TB ports. This creates different trunking opportunities depending on the topology. A two-node connection can dedicate up to six TB ports to the same peer, whereas a four-node fully connected topology leaves at most two trunks for each peer pair. We treat these parallel links as a software trunk so that one logical peer-to-peer transfer can use several physical links. To realize this benefit, the runtime must issue work requests to multiple queue pairs concurrently.

During the setup phase in Fig.~\ref{fig:flow}~(a), the library parses the hostfile used by \textit{JACCL} and expands each peer-pair mapping into one or more RDMA interfaces. For each available TB link, it opens the corresponding RDMA device and creates a queue pair (QP) assigned to a communication worker. We initialize these workers as persistent thread pools rather than creating threads inside every collective call to reduce thread-creation overhead.

At runtime, as shown in Fig.~\ref{fig:flow}~(b), each collective message is partitioned into contiguous 4096-byte chunks\footnote{Unlike conventional InfiniBand devices, Apple's Thunderbolt interface in macOS 26.2 cannot perform hardware-level segmentation, constraining the software-side chunk size to the MTU.}, matching the maximum transmission unit (MTU) of the TB RDMA interface, and statically assigned to the workers associated with the peer pair. Each worker posts send or receive work requests to its own queue pair, so a large logical message is striped across multiple TB links.
For all-reduce operations, the reduction computation runs on the CPU rather than the GPU, avoiding additional GPU scheduling latency for these communication-centric operations that are issued from the CPU. As a result, the performance of all-reduce collectives depends not only on link injection but also on CPU parallelism and memory bandwidth as it helps draw more memory bandwidth through concurrency. This approach preserves the direct-connection communication model of \textit{JACCL} while allowing each peer pair to issue more concurrent RDMA work across the available TB links.

\subsection{Overlapping Gradient Synchronization with Backpropagation}
\begin{figure}[t]
  \centering
  \includegraphics[width=0.9\linewidth]{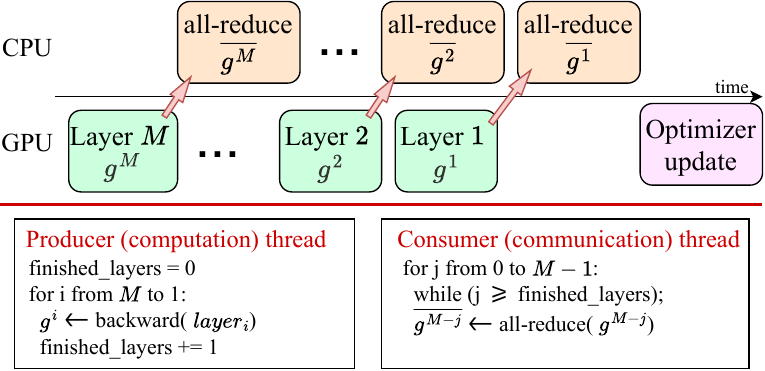}
  \vspace{-1.1em}
  \caption{Layer-wise overlap between GPU backpropagation and CPU-side gradient all-reduce. The producer thread records gradients as layers finish, while the consumer thread launches communication for ready gradients through \textit{JACCL}.}
  \label{fig:overlapping}
  \vspace{-1em}
\end{figure}

Multi-trunk RDMA over TB reduces the time spent in gradient synchronization, but all-reduce remains on the critical path of data-parallel training as it is executed after the  backward pass completes. Apple Silicon provides a useful opportunity for overlap because MLX launches forward and backward computation on the GPU, while \textit{JACCL} coordination and reduction run on CPU threads. We exploit this separation by starting all-reduce for a layer as soon as its gradient becomes available, allowing communication for earlier layers to proceed while the GPU continues backpropagation through later layers.

Fig.~\ref{fig:overlapping} illustrates the layer-wise schedule. A producer thread drives the backward pass and records completed gradient tensors. A consumer thread monitors this queue and invokes the corresponding all-reduce operation when a tensor is ready, using busy-waiting to minimize latency. This consumer is separate from the persistent worker pool used inside the \textit{JACCL} backend, so overlap adds a scheduling thread in addition to the communication workers.

\subsection{Micro-Benchmarks}

We evaluate the multi-trunk RDMA design from Section~\ref{sec::multi-trunk} using micro-benchmarks. Point-to-point send/recv measures the latency and raw bandwidth available to a single peer pair, while all-reduce sum captures the collective primitive used by data-parallel gradient synchronization. The evaluation studies two-node latency and bandwidth across message sizes, examines the effect of software concurrency at 64~MiB, and extends the all-reduce benchmark to larger node counts. Bandwidth results report raw point-to-point bandwidth for send/recv and algorithm bandwidth for all-reduce.

\subsubsection{Two-node Latency}
\label{sec::two-node-latency}
\begin{figure}[t]
    \centering
    \includegraphics[width=.99\linewidth]{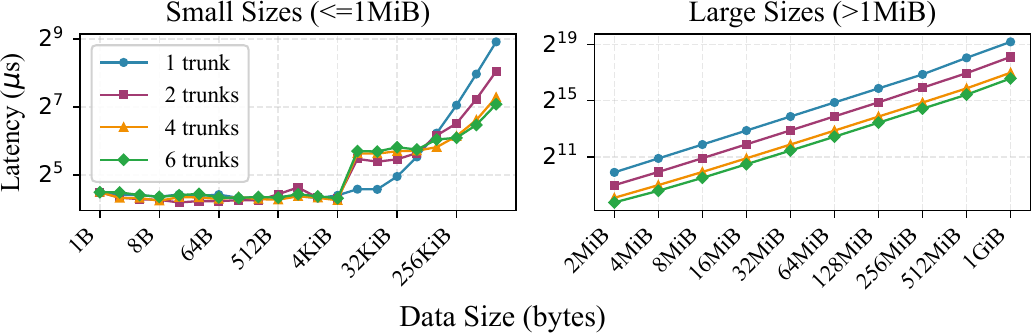}
    \vspace{-1.1em}
    \caption{Two-node send/recv latency across small and large message ranges.}
    \Description{Two side-by-side log-scale line charts of send/recv latency versus data size. The left chart shows messages up to one mebibyte and the right chart shows larger messages. Each chart includes series for one, two, four, and six trunks with all measured data points.}
    \vspace{-0.8em}
    \label{fig:comm-latency-send}
\end{figure}

\begin{figure}[t]
    \centering
    \includegraphics[width=0.99\linewidth]{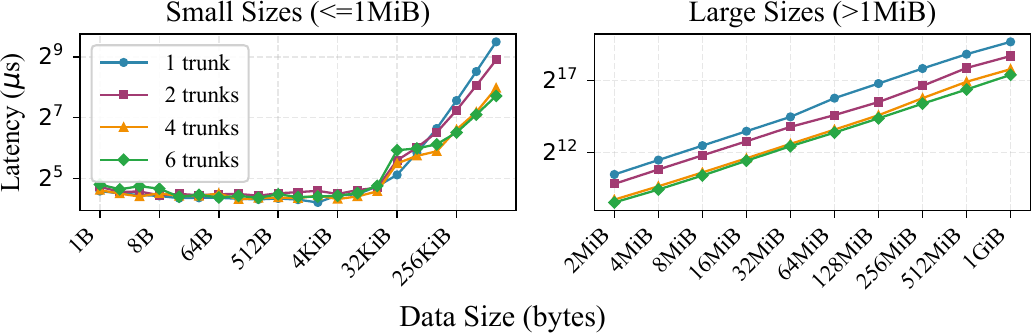}
    \vspace{-1.1em}
    \caption{Two-node all-reduce latency across small and large message ranges.}
    \Description{Two side-by-side log-scale line charts of all-reduce latency versus data size. The left chart shows messages up to one mebibyte and the right chart shows larger messages. Each chart includes series for one, two, four, and six trunks with all measured data points.}
    \label{fig:comm-latency-allreduce}
   
\end{figure}

Fig.~\ref{fig:comm-latency-send} and Fig.~\ref{fig:comm-latency-allreduce} show the latency of send/recv and all-reduce sum, respectively. For small messages, all configurations remain near the software and RDMA overhead floor of roughly 10~$\mu$s. Latency increases noticeably for multi-trunk configurations at 8~KiB, the threshold at which trunking is enabled. Because RDMA devices use a 4~KiB MTU, messages of 8~KiB or larger are split into two or more packets, making them eligible for distribution across multiple trunks. At this size, the threading overhead is still too large to be amortized by the additional link bandwidth.
For messages larger than 1~MiB, latency grows with the number of serialized bytes per link. Using more trunks reduces this serialized component and lowers end-to-end transfer time.

\observation{1}{Multi-trunking is effective for training traffic but not inference traffic.}{Multi-trunking introduces scheduling overhead that cannot be amortized by small messages. Typical tensor-parallel inference activations of 8--16~KiB are below the useful range, while large gradient exchanges from data-parallel fine-tuning can exploit the additional links.}

\subsubsection{Two-node Bandwidth}

\begin{figure}[t]
    \centering
    \includegraphics[width=0.9\linewidth]{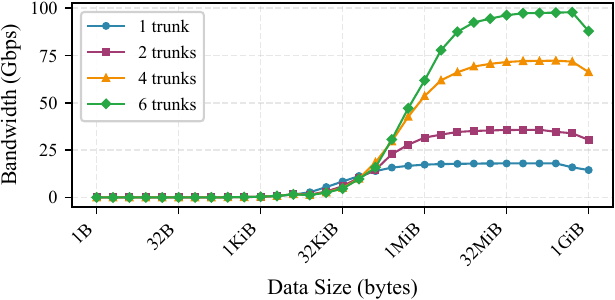}
    \vspace{-1.1em}
    \caption{Two-node send/recv bandwidth across message sizes from 1B to 1GiB.}
    \Description{Line chart of send/recv bandwidth versus message size for one, two, four, and six trunks.}
    \label{fig:comm-bandwidth-send}
\end{figure}

\begin{figure}[t]
    \centering
    \includegraphics[width=0.9\linewidth]{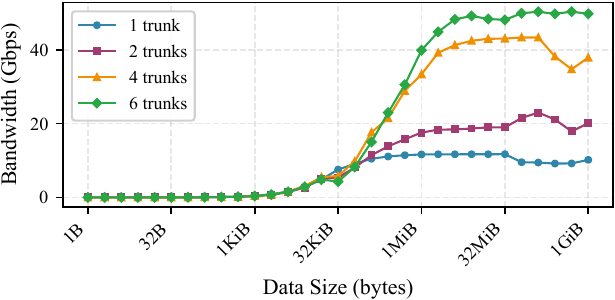}
    \vspace{-1.1em}
    \caption{Two-node all-reduce algorithm bandwidth across message sizes from 1B to 1GiB.}
    \Description{Line chart of all-reduce algorithm bandwidth versus message size for one, two, four, and six trunks.}
    \label{fig:comm-bandwidth-allreduce}
\end{figure}

Fig.~\ref{fig:comm-bandwidth-send} and Fig.~\ref{fig:comm-bandwidth-allreduce} report the bandwidth corresponding to the send/recv and all-reduce sum latency measurements in Section~\ref{sec::two-node-latency}, respectively.
For large messages around 128~MiB, trunking reaches roughly 100~Gbps of raw point-to-point bandwidth with six trunks, compared with about 20~Gbps on one trunk. All-reduce reaches roughly 50~Gbps of algorithm bandwidth, delivering about a \(5\times\) speedup.

\observation{2}{Apple's hardware and kernel RDMA path exposes only a fraction of the nominal TB5 bandwidth.}{A TB5 port is specified for 80~Gbps per direction, but our RDMA path achieves only about 20~Gbps per link. Even with six links, the optimized flow reaches only about 100~Gbps, far below the ideal aggregate bandwidth.}

\subsubsection{Software Concurrency \& Multi-node Scaling}
\begin{figure}[t]
    \centering
    \includegraphics[width=0.99\linewidth]{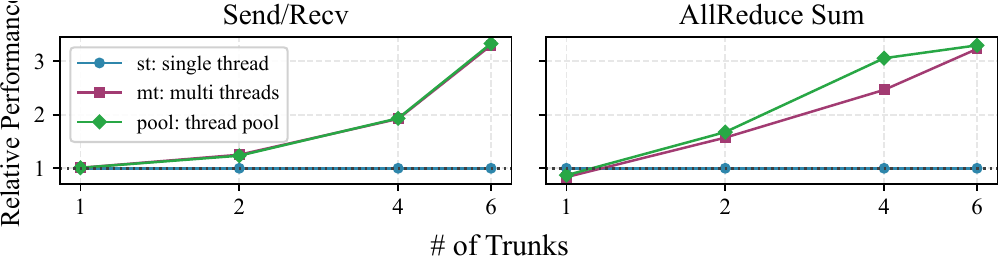}
    \vspace{-1.1em}
    \caption{Software concurrency for 64~MiB two-node communication, normalized to single-thread performance.}
    \Description{Two side-by-side line charts showing relative performance versus trunk count. The left chart is send/recv and the right chart is all-reduce sum. Each chart compares single-thread, multi-thread, and thread-pool implementations normalized to single-thread performance.}
    \label{fig:comm-threading}
\end{figure}

\begin{table}[t]
    \centering
    \caption{All-reduce algorithm bandwidth at 64~MiB across node counts. Values are in Gbps.}
    \vspace{-1.1em}
    \label{tab:comm-scaling}
    \small
    \setlength{\tabcolsep}{4pt}
    \begin{tabular}{llrrrr}
        \toprule
        Nodes & Mode & 1 trunk & 2 trunks & 3 trunks & 6 trunks \\
        \midrule
        2 & st   & 10.7 & 12.5 & 13.8 & 15.6 \\
        2 & pool & 11.0 & 19.0 & 29.1 & \textbf{53.6} \\
        \midrule
        3 & st   & 6.1 & 9.7  & 8.8  & -- \\
        3 & pool & 5.8 & 13.3 & \textbf{21.8} & -- \\
        \midrule
        4 & st   & 5.6 & 5.9  & --   & -- \\
        4 & pool & 5.8 & \textbf{11.6} & --   & -- \\
        \bottomrule
    \end{tabular}
\end{table}

We next examine why the communication runtime needs CPU multithreading and persistent thread pools, as described in Section~\ref{sec::multi-trunk}. Multi-trunk RDMA creates multiple queue pairs per peer pair, but a single CPU thread cannot issue work to all of them fast enough to fully utilize the available links. In addition, all-reduce performs the reduction calculation on the CPU, so collective performance also depends on CPU parallelism and memory bandwidth.

Fig.~\ref{fig:comm-threading} isolates this effect using 64~MiB messages. We compare three implementations. The \textit{st} configuration is single-threaded and uses one communication worker. The \textit{mt} configuration creates multiple worker threads for each communication call. The \textit{pool} configuration uses persistent worker threads initialized during setup, which is the configuration employed in Section~\ref{sec::multi-trunk}.

Normalizing each configuration to the \textit{st} baseline at the same trunk count shows that additional physical links alone are insufficient. A single worker cannot keep multiple queue pairs busy, while multi-threading improves link utilization, and a persistent thread pool provides the same parallelism without paying thread creation costs at every collective. With six trunks, both send/recv and all-reduce improve by more than \(3\times\) relative to the \textit{st} implementation. This result is consistent with the CPU-side reduction design because thread pools expose more CPU parallelism and draw higher memory bandwidth from multiple cores while also driving multiple RDMA queue pairs.

\observation{3}{RDMA-over-TB performance is highly dependent on CPU parallelism.}{Fig.~\ref{fig:comm-threading} shows that CPU multithreading substantially improves both send/recv and all-reduce performance. With six trunks, the multithreaded implementations outperform the single-threaded implementation by more than $3\times$ because additional CPU workers drive multiple queue pairs concurrently and expose more memory bandwidth.}

Finally, Table~\ref{tab:comm-scaling} extends the 64~MiB all-reduce benchmark to larger node counts. The \textit{pool} implementation continues to benefit from additional trunks per peer pair, whereas \textit{st} communication scales only modestly even when more physical links are available. Because each Mac Studio has only six TB ports, the maximum number of trunks per peer pair is three in the three-node
configuration and two in the four-node configuration.

For the \textit{st} configuration, the speedup from using two trunks instead of one decreases from $1.16\times$ to $1.05\times$ as the node count increases from two to four. In contrast, \textit{pool} maintains a similar speedup of around $2\times$. This result shows that all-reduce performance depends heavily on both communication and compute throughput, so both trunking and threading are beneficial.

\subsection{Comparison with 10 Gigabit Ethernet}
\label{sec::tb-vs-10gbe}

\begin{table}[t]
  \centering
  \caption{Communication comparison between 10GbE, original \textit{JACCL}, and our optimized \textit{JACCL}.}
    \vspace{-1.1em}
  \label{tab:tb-10gbe}
  \resizebox{0.99\linewidth}{!}{
  \begin{tabular}{@{}llccc@{}}
    \toprule
    Metric & Nodes & 10GbE & JACCL & \textbf{Our JACCL} \\
    \midrule
    Send/recv latency ($\mu$s) & 2 & 210.9 & 21.3 & \textbf{21.3} \\
    \midrule
    All-reduce bandwidth (Gbps) & 2 & 8.2 & 10.7 & \textbf{53.6} \\
    All-reduce bandwidth (Gbps) & 3 & 6.2 & 6.1 & \textbf{21.8} \\
    All-reduce bandwidth (Gbps) & 4 & 5.5 & 5.6 & \textbf{11.6} \\
    \bottomrule
  \end{tabular}
  }
\end{table}

We also evaluate the Ethernet-based MLX communication backend as a reference point. We use the built-in 10GbE port on the Mac Studio and compare it with the original \textit{JACCL} backend and our optimized \textit{JACCL} backend with multi-trunk communication. Table~\ref{tab:tb-10gbe} reports small-message point-to-point latency and all-reduce algorithm bandwidth at 64~MiB.

In terms of latency, RDMA-over-TB provides about \(10\times\) lower latency than the Ethernet path. Small-message latency is dominated by software and transport overhead, so multi-trunking does not improve this regime. For all-reduce bandwidth, the original \textit{JACCL} is limited by a single trunk and scales similarly to 10GbE at larger node counts. Our optimized \textit{JACCL} uses the maximum available trunk count for each topology, reaching 53.6~Gbps on two nodes, 21.8~Gbps on three nodes, and 11.6~Gbps on four nodes.

This analysis shows that although the original \textit{JACCL} backend released by Apple can provide significantly lower latency for small messages, the bandwidth it can draw still performs similarly to the Ethernet alternative. Our multi-trunking and thread-pool implementation achieves higher bandwidth, which can directly benefit LLM training workflows.

  \section{Fine-Tuning Performance}
\label{sec:sft}

This section evaluates Apple Silicon as a practical platform for private LLM fine-tuning. We study Qwen3-9B fine-tuning from three perspectives. First, we measure the single-node memory and throughput limits to identify which long-context workloads fit within unified memory without context parallelism. Second, we evaluate weak scaling on a four-node Mac Studio cluster and quantify how multi-trunk RDMA-over-TB communication and CPU-side gradient overlap improve aggregate throughput. Finally, we compare the resulting system-level performance and cost tradeoffs with an NVIDIA H100 platform.

\subsection{Experimental Setup}

Our Apple Silicon experiments are conducted on a four-node Mac Studio cluster. Each node is equipped with an M3 Ultra SoC that features 32 CPU cores, 80 GPU cores, and 512~GB of unified memory~\cite{appleMacStudioSpecs}. Nodes are connected using RDMA over Thunderbolt, as described in Section~\ref{sec::multi-trunk}, and use our optimized multi-trunk \textit{JACCL} backend for collective communication. For comparison, we also evaluate an NVIDIA H100 SXM system with 80~GiB of HBM memory and NVLink connectivity~\cite{nvidiaDGXH100UserGuide}.

\begin{table}[t]
  \centering
  \caption{Platform cost and hardware capability comparison. Costs are approximate acquisition costs in US\$.}
    \vspace{-1.1em}
  \label{tab:platform-specs}
  \resizebox{0.99\linewidth}{!}{
  \begin{tabular}{@{}lccc@{}}
    \toprule
    Metric & Apple node & H100 card & 8$\times$H100 DGX \\
    \midrule
    Cost & \$10{,}000 & \$30{,}000 & \$300{,}000 \\
    System scope & Complete node & Single card & Complete node \\
    Theoretical FP32 & 28.3~TFLOP/s & 67.3~TFLOP/s & 538.4~TFLOP/s \\
    Memory bandwidth & 820~GB/s & 3350~GB/s & 2680~GB/s \\
    Memory size & 512~GB & 80~GB & 640~GB \\
    \bottomrule
  \end{tabular}
  }
\end{table}

Table~\ref{tab:platform-specs} summarizes the cost and hardware capability tradeoff between the Apple Silicon node used in our cluster and H100-based configurations. The Apple Silicon node has lower theoretical FP32 throughput and memory bandwidth than an H100 accelerator, but it provides substantially larger memory capacity per complete node at a lower acquisition cost.

We benchmark supervised fine-tuning of Qwen3-9B~\cite{yang2025qwen3technicalreport} using the SGD optimizer. The benchmark follows the structure of the Hugging Face model implementation and is ported to Apple Silicon using MLX and our customized \textit{JACCL} collective backend. We use mini-batch data-parallel training, where each node computes gradients on a local mini-batch before synchronizing gradients with all-reduce. The scalability experiments are conducted through weak scaling, keeping the local mini-batch size fixed on each node while increasing the global mini-batch size with the number of nodes. This setting reflects common fine-tuning workflows that process many mini-batches and provides a direct measure of end-to-end system throughput~\cite{shoeybi2019megatron,2025llama3}.

We evaluate two optimizations introduced in Section~\ref{sec::multi-trunk}. Multi-trunk communication uses multiple RDMA-over-TB links for each peer pair when the topology provides more than one direct path. CPU-side gradient overlap starts all-reduce operations during backpropagation, allowing communication to be hidden behind remaining GPU computation when sufficient backward work remains.

\subsection{Single-Node Limits}

\begin{figure}[t]
  \centering
  \includegraphics[width=0.95\linewidth]{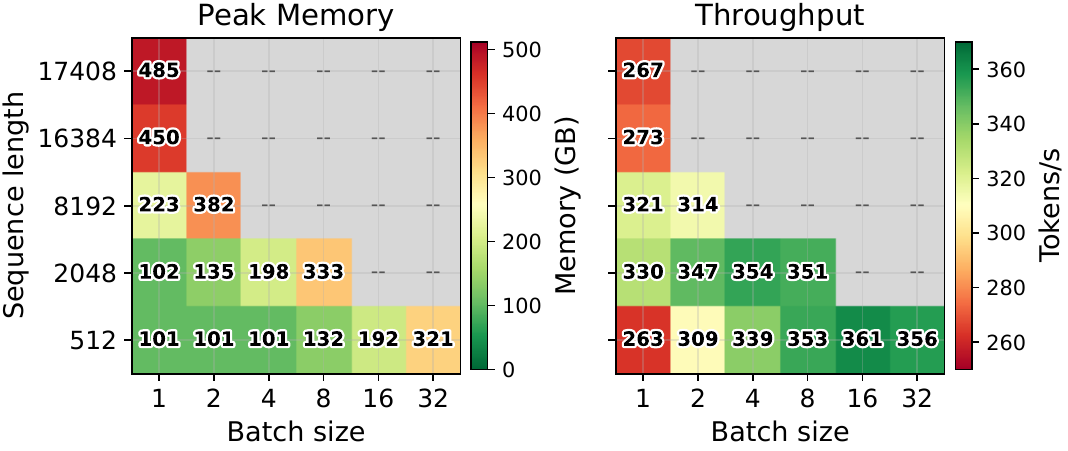}
    \vspace{-1.1em}
  \caption{Single-node memory footprint and training throughput for Qwen3-9B supervised fine-tuning across sequence lengths and batch sizes.}
  \Description{Two heatmaps. The first shows peak memory by sequence length and batch size. The second shows single-node tokens per second across the same settings.}
  \label{fig:sft-single-node}
\end{figure}

Fig.~\ref{fig:sft-single-node} shows heatmaps of single-node memory use and training throughput across sequence lengths and batch sizes. The unified-memory capacity allows Qwen3-9B fine-tuning at substantially larger sequence lengths than an 80~GB accelerator can support. The largest evaluated sequence length, 17408, fits within a single Apple Silicon node without employing context parallelism. These results show that the maximum evaluated context length can be served on one node, while memory capacity determines which batch-size configurations remain feasible.

At a fixed sequence length, larger batch sizes generally provide higher throughput because they increase arithmetic intensity and amortize optimizer-update overhead over more tokens. Throughput should still be interpreted within each sequence-length setting. The computational complexity of the attention block scales quadratically with context length, so longer contexts perform more work per token and naturally achieve lower tokens/s.

\subsection{Cluster Weak Scaling}

\begin{table}[t]
  \centering
  \caption{Weak-scaling throughput on the Mac Studio cluster for sequence length 17408 and local batch size 1. Each entry reports tokens/s followed by wall time in seconds.}
    \vspace{-1.1em}
  \label{tab:sft-weak-scaling}
  \resizebox{0.99\linewidth}{!}{
  \begin{tabular}{@{}ccllll@{}}
    \toprule
    Nodes & Mode & 1 trunk & 2 trunks & 3 trunks & 6 trunks \\
    \midrule
    1 & -- & 264 / 66 & -- & -- & -- \\
    \multirow{2}{*}{2} & w/o overlap & 348 / 100 & 431 / 81 & 452 / 77 & 490 / 71 \\
      & w/ overlap & 472 / 74 & 484 / 72 & 488 / 71 & \textbf{492 / 71} \\
    \multirow{2}{*}{3} & w/o overlap & 479 / 109 & 592 / 88 & 654 / 80 & -- \\
      & w/ overlap & 684 / 76 & 713 / 73 & \textbf{717 / 73} & -- \\
    \multirow{2}{*}{4} & w/o overlap & 571 / 122 & 736 / 95 & -- & -- \\
      & w/ overlap & 776 / 90 & \textbf{936 / 74} & -- & -- \\
    \bottomrule
  \end{tabular}
  }
\end{table}

\begin{figure}[t]
  \centering
  \includegraphics[width=0.7\linewidth]{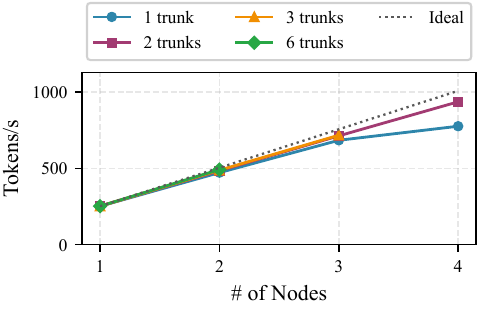}
    \vspace{-1.1em}
  \caption{Weak-scaling throughput on the Mac Studio cluster for sequence length 17408 and local batch size 1 across different trunk counts.}
  \Description{Line chart showing tokens per second rising from one to four nodes for one, two, three, and six trunk configurations.}
  \label{fig:sft-weak-scaling}
\end{figure}

\begin{figure}[t]
  \centering
  \includegraphics[width=0.99\linewidth]{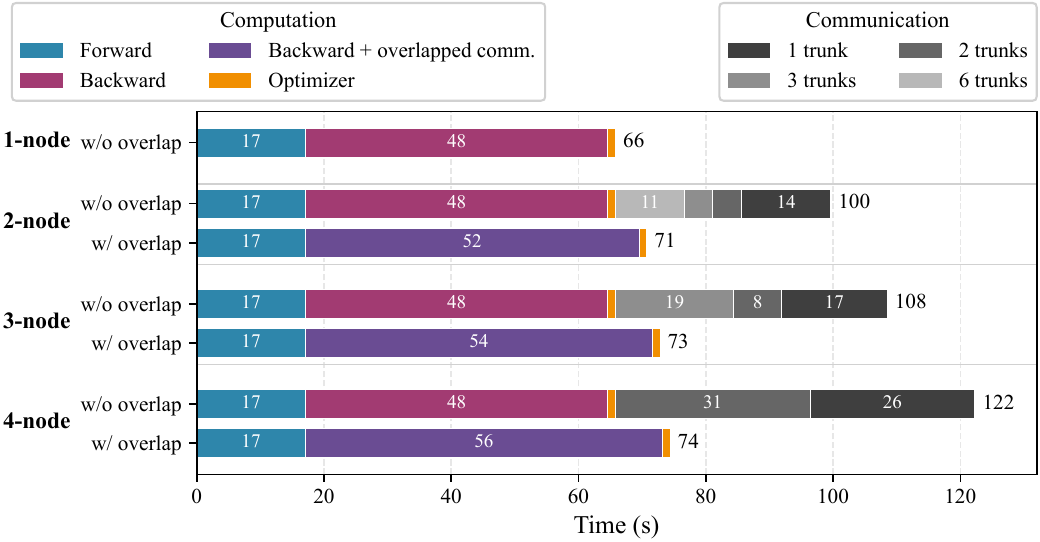}
    \vspace{-1.1em}
  \caption{Iteration-time breakdown for weak-scaling runs with and without CPU-side gradient communication overlap. Non-overlapped bars report exposed communication time by trunk count, while overlapped bars use the maximum available trunk count.}
  \Description{Horizontal stacked bar chart showing forward, backward, communication, and optimizer update time for one through four Mac Studio nodes.}
  \label{fig:sft-overlap-breakdown}
\end{figure}

Weak scaling is the most suitable evaluation mode for LLM training because the training set is much larger than the local batch processed by each node. Practical fine-tuning workloads contain thousands or more training samples, so adding nodes usually increases the number of samples processed per iteration rather than splitting a fixed small batch across more devices. As a result, we keep the local batch size fixed on each node and evaluate the aggregate throughput achieved by the whole system.

Table~\ref{tab:sft-weak-scaling} reports the weak-scaling sweep with sequence length 17408 and local batch size 1. The \textit{w/o-overlap} rows provide two baselines. The 1 trunk entries show performance without multi-trunking, while the trunk counts above 1 show the benefit of multi-trunking without communication overlap. Increasing the trunk count improves end-to-end throughput whenever additional direct links are available. The \textit{w/ overlap} rows then add CPU-side gradient communication overlap on top of multi-trunk communication. With the maximum available trunk count for each topology, the optimized two-node, three-node, and four-node runs retain 93\%, 90\%, and 89\% of the single-node efficiency, respectively, computed as one-node time divided by multi-node time. Fig.~\ref{fig:sft-weak-scaling} plots the corresponding aggregate throughput trend, using the one-node result as the shared no-communication baseline. The best four-node result reaches 936 tokens/s, which is 3.5$\times$ the one-node throughput.

Fig.~\ref{fig:sft-overlap-breakdown} further breaks down iteration time with and without communication overlap. For non-overlapped runs, it separates communication time by trunk count to show how exposed synchronization cost is distributed across links. Communication grows with the number of nodes and becomes a major part of iteration time without overlap. Multi-trunking reduces this exposed cost, while overlap hides more CPU-side gradient communication behind the GPU-heavy backward pass. In the four-node case, overlap reduces iteration time from 122~s to 74~s, giving a 1.6$\times$ speedup.

\observation{4}{Apple Silicon's CPU--GPU execution separation enables communication overlap.}{RDMA work submission and all-reduce reduction execute on CPU threads, while forward and backward computation executes on the GPU. This separation allows gradient synchronization to overlap with backpropagation, hiding communication that multi-trunking alone cannot eliminate.}

\subsection{Platform Comparison}

\begin{figure}[t]
  \centering
  \includegraphics[width=0.8\linewidth]{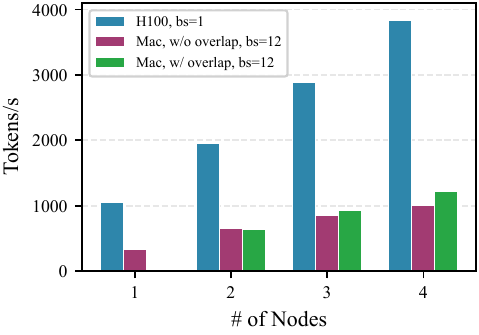}
    \vspace{-1.1em}
  \caption{Weak-scaling throughput comparison between H100 and Mac Studio systems at sequence length 2048. The H100 run uses local batch size 1, while the Mac Studio run uses local batch size 12.}
  \Description{Grouped bar chart comparing H100 throughput, Mac throughput without overlap, and Mac throughput with overlap as node count increases.}
  \label{fig:sft-platform-comparison}
\end{figure}

To evaluate whether Apple Silicon is competitive with an NVIDIA counterpart, we use sequence length 2048, which fits within the NVIDIA H100's 80~GB memory. We run the same workload on both platforms and choose the local batch size that fits each system's memory capacity. The resulting throughput is shown in Fig.~\ref{fig:sft-platform-comparison}.

The H100 delivers higher raw throughput, reaching 1045 tokens/s on one node and 3830 tokens/s on four nodes with local batch size 1, while the Mac Studio cluster reaches 1210 tokens/s on four nodes with overlapped communication and local batch size 12. This result shows that four Mac Studio nodes can exceed the throughput of one H100 accelerator on this workload.

Table~\ref{tab:platform-specs} lists the approximate acquisition cost of a complete Apple Silicon node at \$10{,}000 and a single H100 accelerator at \$30{,}000. Although four Mac Studio nodes cost more than a single H100 card, Apple Silicon provides two practical advantages in our setting. First, the H100 has much smaller memory capacity, which limits the maximum sequence length and batch configurations that can be trained without additional parallelism. Second, a common H100 system such as an 8$\times$H100 DGX costs around \$300{,}000 and provides 640~GB of aggregate HBM memory, while four Apple Silicon nodes provide up to 2~TB of aggregate unified memory.

Memory capacity determines which workloads can execute, while computational throughput determines how quickly a feasible workload finishes. For small organizations that cannot maintain server-grade GPU systems, Apple Silicon offers a lower-cost path to fine-tuning with greater capacity for long training samples.

  \section{Conclusion}
\label{sec:conclusion}

This paper studied Apple Silicon as a platform for private LLM fine-tuning, where memory capacity determines feasibility and communication efficiency determines multi-node scalability. We extended \textit{JACCL} with multi-trunk communication and persistent workers, and combined these mechanisms with CPU-side gradient overlap for MLX training.

Our evaluation highlights four properties of RDMA-over-TB. First, multi-trunking improves throughput for large data-parallel gradient exchanges but offers little advantage for the 8--16~KiB per-token activation transfers typical of distributed inference. Second, Apple's hardware and kernel path exposes only a fraction of nominal TB5 bandwidth, delivering about 20~Gbps per link and 100~Gbps across six links. Third, communication performance also depends strongly on CPU parallelism, with six-trunk multithreading exceeding single-threaded performance by more than \(3\times\). Finally, CPU-side communication can overlap with GPU computation. At four nodes with two trunks, overlap reduces iteration time from 95~s to 74~s and increases throughput from 736 to 936 tokens/s.

Apple Silicon's unified memory also supports LLM fine-tuning with longer samples than a single H100 can accommodate without context parallelism. On four Mac Studios, multi-trunk communication and overlap improve weak-scaling throughput by up to \(1.6\times\) over the single-trunk, non-overlapped baseline. Although H100 offers higher raw throughput, Apple Silicon provides greater memory capacity per complete node at lower acquisition cost, making it a practical platform for private, memory-constrained fine-tuning.

  \begin{acks}
We acknowledge financial support from Academia Sinica's SiliconMind Project (AS-IAIA-114-M11) and the National Science and Technology Council (NSTC) of Taiwan under grant no. 115-2221-E-002-149-MY3. Additional support came from the Center of Data Intelligence: Technologies, Applications, and Systems, National Taiwan University, under the Featured Areas Research Center Program within the framework of the Higher Education Sprout Project by the Ministry of Education of Taiwan (grant no. NTU-115L900903). We also thank the National Center for High-performance Computing and Taipei-1 for providing computational resources.
\end{acks}

  \bibliographystyle{ACM-Reference-Format}
  \bibliography{refs.bib}

  \end{document}